\newcommand{\cv}[1]{}
\newcommand{\av}[1]{#1}
\pdfoutput=1
\documentclass[letterpaper]{article}
\usepackage{geometry}
\usepackage[T1]{fontenc}
\usepackage[utf8]{inputenc}
\usepackage{newtxtext}
\usepackage{helvet}
\usepackage{courier}
\usepackage{microtype}
\DisableLigatures[i,I]{encoding=T1,family=*}  %
\usepackage{amsmath,amssymb,amsthm}
\usepackage{graphicx}
\usepackage{xcolor}
\usepackage[authoryear]{natbib}
\bibpunct{[}{]}{;}{a}{}{,}
\usepackage[hyphens]{url}
\usepackage[hidelinks]{hyperref}
\hypersetup{%
  pdftitle={Streaming LRAT Certificates into Lean Theorems},%
  pdfauthor={Stefan Szeider},%
  pdfkeywords={SAT solving, LRAT certificates, Lean 4, proof by reflection, cube-and-conquer, verified checking}}
\date{}
\makeatletter
\let\@dblfloat\@float
\let\end@dblfloat\end@float
\makeatother
\theoremstyle{definition}
\newtheorem{example}{Example}
\newcommand{\modesFirst}{2.3cm}\newcommand{\modesCol}{1.42cm}\newcommand{\modesSep}{3pt}
\newcommand{\injFails}{\raggedright\arraybackslash}\newcommand{\casesA}{1.7cm}\newcommand{\casesB}{3.0cm}\newcommand{\casesC}{1.3cm}\newcommand{\casesD}{1.4cm}\newcommand{\casesE}{1.4cm}\newcommand{\casesSep}{2pt}

\usepackage{array}
\usepackage{booktabs}
\usepackage{multirow}
\usepackage{listings}
\usepackage{needspace} %
\usepackage{xspace}
\usepackage{tikz}
\usetikzlibrary{positioning,fit}
\usepackage{cleveref}

\newcommand{\tightlines}{\setlength{\baselineskip}{0.82\baselineskip}}

\newcommand{\tool}{lrat-catcher}

\cv{\AtBeginDocument{}}

\lstdefinelanguage{lean}{
  morekeywords={theorem,def,lemma,example,where,let,in,if,then,else,match,with,
    do,return,have,show,by,sorry,import,open,namespace,end,section,variable,
    instance,class,structure,inductive,noncomputable,deriving,extends,abbrev,
    axiom,private,protected,mutual,attribute,macro,syntax,notation,Type,Prop,
    Sort,true,false,native_decide,decide},
  sensitive=true,
  morecomment=[l]{--},
  morecomment=[s]{/-}{-/},
  morestring=[b]",
}
\av{\lstset{xleftmargin=0.1\textwidth,xrightmargin=0.1\textwidth}}
\definecolor{leanblue}{rgb}{0,0,0.55}
\definecolor{leangreen}{rgb}{0,0.40,0}
\colorlet{leancode}{leanblue}
\newcommand{\leanfontsize}{\small}
\lstdefinestyle{inl}{basicstyle=\leanfontsize\ttfamily\color{leancode},keywordstyle=\bfseries\color{leancode}}
\usepackage{etoolbox}
\makeatletter\av{\g@addto@macro\@floatboxreset{\centering}}\makeatother
\av{\AddToHook{env/table/begin}{\setlength{\belowcaptionskip}{\medskipamount}}\AddToHook{env/table*/begin}{\setlength{\belowcaptionskip}{\medskipamount}}}
\AtBeginEnvironment{table}{\renewcommand{\leanfontsize}{}}
\AtBeginEnvironment{table*}{\renewcommand{\leanfontsize}{}}
\newcommand{\lean}{\lstinline[style=inl,breaklines=true,breakatwhitespace=true]}

\newcommand{\injectionstable}{%
\begin{table}
\caption{Fault injections on a six-chunk file-mode PHP(10,9) import
and on a stream-mode PHP(9,8) import
(\Cref{sec:evalT2}). Outcome
is what Lean does; ``check false'' = the mutated object reaches a
step check that evaluates to false, stopping the
build.}
\label{tab:injections}
\footnotesize
\setlength{\tabcolsep}{3pt}
\renewcommand{\arraystretch}{1.1}
\begin{tabular}{@{}>{\tightlines\raggedright\arraybackslash}p{4.2cm}>{\tightlines\injFails}p{1.6cm}>{\tightlines\raggedright\arraybackslash}p{1.7cm}@{}}
\toprule
injection & fails at & outcome\\
\midrule
none (control) & --- & proved, 38 s\\
snapshot of a different formula & chunk 2 & check false\\
earlier snapshot of the same import & chunk 2 & check false\\
one digit changed in a snapshot clause & chunk 2 & check false\\
snapshot hole replaced by a clause & chunk 2 & check false\\
chunk certificate replayed as the next & chunk 3 & check false\\
chunk skipped (resume from earlier snapshot) & chunk 3 & check false\\
two chunk certificates swapped & chunk 2 & check false\\
final chunk truncated (no empty clause) & chunk 6 & check false\\
snapshot truncated mid-line & chunk 2 & parses to garbage, check false\\
one hint id changed in a chunk & chunk 2 & check false\\
artifacts of chunks 3--6 deleted & --- & resumed, 7 s, same axioms\\
\addlinespace
stream: certificate truncated at 60\% & block 10, end of stream & check false\\
stream: garbage bytes mid-certificate & block 7, parser & check false\\
stream: producer killed at 43\% & block 6, parser & check false\\
\bottomrule
\end{tabular}
\end{table}
}

\begin{document}

\title{Streaming LRAT Certificates into Lean Theorems}

\av{\author{Stefan Szeider\\[4pt]
  \small Algorithms and Complexity Group\\[-2pt]
  \small TU Wien, Vienna, Austria\\[-2pt]
  \small \texttt{sz@ac.tuwien.ac.at}\\[-2pt]
  \small \href{https://www.ac.tuwien.ac.at/people/szeider/}{https://www.ac.tuwien.ac.at/people/szeider/}}
\maketitle
\thispagestyle{empty}}
\begin{abstract}
If the certificate produced by a SAT solver is checked by a
verified checker, we get a verdict which convinces. But this verdict
cannot be named, reused as a lemma, or composed with other formal
developments. We propose the tool \tool{}, which turns a certificate
into a Lean theorem. It checks the certificate as a stream while the
solver is still running. Hence the certificate is not required to be
saved to a file. Additionally, our tool makes Lean core's verified
LRAT checker resumable so that its state can be serialized. We prove
that checking divided at such a state still properly refutes the
original formula. We propose two import modes. The stream mode reads
the certificate from a pipe in blocks and checks it on the fly in
memory. The file mode imports a stored certificate in chunks. If
interrupted, it rechecks only the chunks it has not yet completed. The
soundness theorem for the stream mode guarantees that a garbled or
adversarial stream can only fail the check but not yield a false
theorem. We find that with compaction at chunk boundaries, the memory
required depends only on the live clause set, not on the certificate size. Our tool supports cube-and-conquer and formulas derived by
preprocessing, to still form Lean proofs of the original formula. We
provide several end-to-end case studies on well-known
combinatorial problems. A larger scaling experiment on the
empty-hexagon problem shows that 174~TB of certificates can be
imported into Lean via streaming.
\end{abstract}

\cv{\maketitle}

\section{Introduction}\label{sec:intro}                           %

When we solve a combinatorial question with a SAT solver, we receive
the answer in terms of a verdict. With a verified checker of the
solver's LRAT certificate \citep{LRAT}, we can make the verdict more
convincing. Inside the proof assistant, the unsatisfiability of the formula is not just a verdict, but it is a theorem that references exactly the refuted formula. So the fact that the formula is unsatisfiable becomes a theorem, a Lean object with a name. The fact that was proved is a Lean statement, not the run of a tool. The theorem can be named, reused, and composed; in
contrast, a verdict cannot. When the encoding of the combinatorial
question into SAT is accomplished within Lean itself, the correctness
of this encoding can be proved in Lean. This way we get a pipeline
that runs end to end from the mathematical statement to the refuted
formula. This paper is about the step from the certificate to the
theorem in Lean~4 \citep{Lean4}, at the scale of current SAT
computations, which reach certificates of hundreds of terabytes.

We describe a tool, \tool{}\av{\footnote{\url{https://github.com/leansolving/lrat-catcher}}}, which turns a SAT solver's LRAT
certificate into a Lean theorem by checking the certificate received
from the solver as a stream \emph{while it is running}. Its input is
a CNF formula, given as a DIMACS file or as a Lean term that has been
produced by an encoding written in Lean, and an LRAT certificate
received from any solver as a stream or as a file. The tool is
reusable and not specific to a certain problem or encoding.

With cube-and-conquer, one divides the computation on hard
instances into smaller pieces. However, this does not make the
certificates small. The split is chosen for the solver, not for the
checker. In the computations we deal with, the certificates of
single cubes can range from kilobytes to more than ten gigabytes.
If we split large cubes further, we end up with possibly many more
cubes and interfere significantly with the solving process itself.
Another issue is that the certificate of a preprocessing step applied
to the original formula, or a certificate produced by others, is not
cubed at all.

A verified checker that reads a certificate as it is produced must
proceed piece by piece and form one sound computation. And since a
certificate that is not kept cannot be re-read, its soundness theorem
must cover any possible input.

\cv{\textbf{(1)} }\av{\paragraph{(1)}}Our tool makes the verified LRAT checker of Lean core
  \citep{Boving25} \emph{resumable} (\Cref{sec:checker}). We divide a check
  into the checks of chunks. Each starts from a checker state that
  has been produced by the previous chunk and comes with a soundness
  lemma of its own. The chain of passed checks is proved to refute
  the original formula. This construction uses the core checker's
  public per-step lemmas and adds no trusted code.

\cv{\textbf{(2)} }\av{\paragraph{(2)}}Our tool has two import modes (\Cref{sec:streaming}). In
  \emph{stream mode}, the Lean process reads the certificate from a
  pipe in blocks. It runs the checks block by block in memory, and
  once the empty clause has been derived, it elaborates the theorem
  that the input formula is unsatisfiable. This way, no certificate
  byte is ever written to \emph{persistent storage}, i.e., the
  filesystem. The soundness theorem quantifies over all possible
  streams. The only trusted component that is added by the stream
  mode is the code that reads the pipe and supplies the streamed
  bytes to the checker. In \emph{file mode}, we divide a stored
  certificate into chunks. Each chunk is checked in a compiled Lean
  file of its own, starting from the serialized state of the
  previous chunk. The file keeps the chunk's text so that after an
  interruption, only the chunks without a compiled file need to be
  checked again. We use a generator to write these files.

\cv{\textbf{(3)} }\av{\paragraph{(3)}}Our tool supports \emph{cube-and-conquer} \citep{CubeAndConquer}
  (\Cref{sec:compose}). It imports the leaves of a split in parallel
  as independent Lean modules. One lemma combines their theorems with
  the cover theorem into one theorem about the full formula. We
  compose certified derivations, which transform a formula instead
  of refuting it, the same way, so that the theorem is about the
  original formula.

\cv{\textbf{(4)} }\av{\paragraph{(4)}}We \emph{evaluate} our tool and compare it with verified and
  unverified external checkers (\Cref{sec:eval}). It imports a
  certificate as a theorem and requires about 15 times the CPU time
  of lrat\_isa \citep{Lammich24}, the fastest verified checker we
  considered, on a 276~MB book-graph Ramsey certificate
  (\Cref{sec:evalT1}). The required memory is bounded in terms of the
  number of live clauses and not the certificate size. This is achieved by compacting the checker state at boundaries and removing the entries of deleted clauses. This required less than 4~GB on the certificate of around 500~GB.

\cv{\textbf{(5)} }\av{\paragraph{(5)}}We test the reusability of our tool by running it on
  \emph{five end-to-end case studies} (\Cref{sec:evalT3}). Each is a named
  theorem whose proof involves a SAT encoding written in Lean and a
  cube-and-conquer import. Our case studies include the Ramsey number $R(4,4) = 18$, the queen-domination enumeration for boards up to $19 \times 19$, confirming the numbers by \citet{RostamiBright25}, who corrected the wrong number reported in OEIS A002563 \citep{OEIS-A002563}, the fact that the Keller graph $G_{7,3}$ is clique-free, and the van der Waerden numbers $w(2;3,17) = 279$ and $w(2;3,18) = 312$. In addition to the case
  studies, we re-solve the empty-hexagon problem (\Cref{sec:evalHex})
  and import its 174~TB of certificates in stream mode without
  storing a single byte.
\av{\bigskip}

\cv{\vspace{6pt}}
For machine-checked theorems that require storing certificates at
terabyte scale, the certificates are often not distributed or made
publicly available. For instance, \citet{BPT16} report almost 200~TB
for the DRAT proof of the Boolean Pythagorean triples theorem; the project page is live, but its link to the 68~GB archive of cube files was dead as of August 2026. \citet{Subercaseaux24} did not store any of the
certificates of the empty-hexagon formalization, so a skeptical
reader must repeat the whole computation. A streamed import as
provided by our tool reduces such a repetition to the solving
itself. The re-solving and rechecking do not require terabyte-scale
storage and, with compaction, not even memory above the set of live
clauses.
\av{We compare our import modes with five other mechanisms in
\Cref{sec:modes}, in terms of what the mechanism keeps and whether
the prover sees a theorem or an axiom.}

\section{Background and Trust}\label{sec:background}             %

\subsection{Clausal certificates}\label{sec:certificates}

A SAT solver receives a propositional formula $F$ in conjunctive
normal form (CNF) and decides whether it is satisfiable or not. The
formula is usually represented in the DIMACS file format, one clause
per line of signed integers. The solver outputs a satisfying
assignment as a witness for satisfiability, or a \emph{clausal
certificate} of unsatisfiability. The latter is a
sequence of learned clauses, each redundant
(satisfiability-preserving) with respect to the preceding clauses, and ends with the
empty clause. \citet{DRATtrim} introduced DRAT as a standard format for clausal
certificates, and \citet{GRIT17} added clause ids and unit-propagation hints in GRIT (RUP only). \citet{LRAT} extended DRAT with ids and hints. Ids and hints make checking easier. \citet{cakelpr} verified the checker cake\_lpr in HOL4. With LRAT, the
certificate can be checked on the fly while the solver is running,
as CaDiCaL writes it directly \citep{Pollitt23}. A solver that
writes DRAT needs an LRAT conversion from a stored file.

An LRAT certificate is a sequence of \emph{actions}, written one per
line in the textual format. Actions can be \emph{addition} lines,
which give a fresh clause id, the literals of the new clause, and,
after a zero, the \emph{hints}, which are ids of earlier clauses; or
\emph{deletion} lines, which give a label, the marker \lean|d|, and
the ids of the clauses to remove. Addition lines represent \emph{reverse unit propagation} (RUP), where
a new clause is implied by hint clauses through unit propagation.
The addition can be checked by assuming the negation of the literals of the new clause,
where, in hint order, each hint clause but the last becomes unit
under the accumulated assumptions and adds its forced literal to
them, and the last hint clause is falsified. LRAT also allows \emph{RAT} additions (resolution
asymmetric tautology), which preserve satisfiability without being
implied and are used for transformations such as symmetry breaking. Consequently, additions preserve satisfiability, and
deletions only remove constraints. Thus, if $F$ were satisfiable,
every clause set along the certificate would remain satisfiable and
could not contain the empty clause. A certificate that contains the empty
clause \emph{refutes} $F$.

\begin{example}\label{ex:running}
Consider as a running example the following formula $F$ over the
variables $x_1, x_2, x_3$; clauses are numbered in file order:
{\allowdisplaybreaks
\begin{alignat*}{2}
C_1 &= x_1 \lor x_2 \lor x_3, &\quad
C_2 &= x_1 \lor x_2 \lor \lnot x_3,\\
C_3 &= x_1 \lor \lnot x_2, &
C_4 &= \lnot x_1 \lor x_2,\\
C_5 &= \lnot x_1 \lor \lnot x_2 \lor x_3, &
C_6 &= \lnot x_1 \lor \lnot x_2 \lor \lnot x_3.
\end{alignat*}}
Its LRAT refutation has five additions and two deletions,
annotated after \lean|--|:
\av{\Needspace*{9\baselineskip}}
\begin{lstlisting}
7 1 2 0  1 2 0    -- add 7: x₁ ∨ x₂   (hints 1, 2)
8 1 0  7 3 0      -- add 8: x₁        (hints 7, 3)
8 d 2 0           -- delete clause 2
9 -1 -2 0  5 6 0  -- add 9: ¬x₁ ∨ ¬x₂ (hints 5, 6)
9 d 7 0           -- delete clause 7
10 -1 0  9 4 0    -- add 10: ¬x₁      (hints 9, 4)
11 0  8 10 0      -- add 11: empty    (hints 8, 10)
\end{lstlisting}
The second action assumes $\lnot x_1$, which makes hint 7 force
$x_2$, and falsifies hint 3; the final action derives the empty
clause from the unit clauses 8 and 10.
\end{example}

\noindent
All refutation certificates considered in this paper are RUP-only
(CaDiCaL produces no RAT step with factoring off, \Cref{sec:eval});
the certified derivations (\Cref{sec:deriv}), however, use RAT
throughout.

\subsection{From check to theorem by reflection}\label{sec:reflection}

An \emph{import} of a certificate for a formula $F$ is a Lean
declaration that proves \lean|CNF.Unsat| of the \lean|CNF Nat| term
that denotes $F$. Here, the term that denotes $F$ is an array of
clauses over natural-number variables. \lean|CNF.Unsat| states that
it has no satisfying assignment. With \emph{reflection}, we obtain
such a declaration from running a verified checker on $F$ and the
certificate. We use the formally verified LRAT checker
\citep{Boving25} of Lean core \citep{Lean4} (the Lean distribution
with its standard library, not the trusted kernel), which is used by
Lean's \lean|bv_decide| tactic. It takes $F$ and a complete parsed
certificate and returns \lean|true| if the certificate is valid. Its
soundness lemma \lean|check_sound| translates this result into
\lean|.Unsat| of $F$. Internally, the checker keeps a clause
database, which is indexed by clause ids, and each action step has a
public soundness lemma, which we will use directly
(\Cref{sec:checker}).

Our function \lean|checkLrat| takes the DIMACS file and the LRAT
certificate as strings (\lean|cnfStr|, \lean|lratStr|), parses both
(the formula with \lean|parseDimacs|), and runs the core checker.
The parser is kept short because a fault in it changes which formula
the theorem is about (\Cref{sec:trust}). We refer to a result that is
proved once and for all and is part of our library as a
\emph{lemma}; see \Cref{tab:lemmas} for an overview of our lemmas.
In contrast, an import creates \emph{theorems} that are specific
to the considered formula.
\begin{table}
\caption{The lemmas of the library, 23 declarations. Each lemma is introduced in the section listed; a second name in a row is a
variant, over a CNF term or for the middle step.}
\label{tab:lemmas}
\footnotesize
\setlength{\tabcolsep}{4pt}
\begin{tabular}{@{}llr@{}}
\toprule
Lemma & Lean name & \llap{Section}\\
\midrule
1 & \lean|checkLrat_sound|, \lean|checkLratCnf_sound| & \ref{sec:reflection}\\
2 & \lean|stepStart_sound| & \ref{sec:chaining}\\
3 & \lean|stepMid_sound| & \ref{sec:chaining}\\
4 & \lean|stepFinish_sound| & \ref{sec:chaining}\\
5 & \lean|runChunk_refuted_sound| & \ref{sec:runchunk}\\
6 & \lean|runChunk_more_sound| & \ref{sec:runchunk}\\
7 & \lean|restoreD_serialize| & \ref{sec:runchunk}\\
8 & \lean|restoreD_ready| & \ref{sec:runchunk}\\
9 & \lean|checkStream_sound|, \lean|checkStreamCnf_sound| & \ref{sec:streammode}\\
10 & \lean|checkStream_feed_agree|, \lean|checkStreamCnf_feed_agree| & \ref{sec:streammode}\\
11 & \lean|units_empty_of_ready| & \ref{sec:compaction}\\
12 & \lean|liff_restoreD_compactSnapshot_serialize| & \ref{sec:compaction}\\
13 & \lean|checkStreamC_sound|, \lean|checkStreamCnfC_sound| & \ref{sec:compaction}\\
14 & \lean|stepStart_restores|, \lean|stepMid_restores| & \ref{sec:deriv}\\
15 & \lean|extern_unsat_restoreD| & \ref{sec:deriv}\\
16 & \lean|stepStart_extern|, \lean|stepMid_extern| & \ref{sec:deriv}\\
17 & \lean|cover_unsat| & \ref{sec:cover}\\
\bottomrule
\end{tabular}
\end{table}
Lemma~1 is the soundness lemma of \lean|checkLrat|. It states that
\lean|checkLrat cnfStr lratStr = true| implies
\lean|(parseDimacs cnfStr).Unsat|, for all strings \lean|cnfStr|
and \lean|lratStr|. Here, the certificate appears only in the
hypothesis. Hence, it is not trusted, and a failed parse makes the
check fail. The term that denotes $F$ is \lean|(parseDimacs cnfStr)|,
where \lean|cnfStr| is the DIMACS text of $F$ embedded as a string
literal in the declaration's module. Alternatively, it can be the
term produced by an encoding that is written in Lean; the latter is
used in the end-to-end pipeline (\Cref{sec:trust}).

If we replay the certificate inside the kernel, one proof step per
action, we obtain a proof term whose size grows with the
certificate. Mathlib's \lean|lrat_proof| command \citep{FromLRAT}
does this for RUP-only certificates; we will see later where this
approach runs out of memory (\Cref{sec:evalT1}). Reflection,
however, proves the declaration with the single term
\lean|checkLrat_sound cnfStr lratStr h|, where \lean|h| is a proof
of \lean|checkLrat cnfStr lratStr = true|. \emph{Elaboration} is the
process performed by Lean when it processes a declaration and
produces the term, which is checked by the Lean kernel. An import is
elaborated once, at the time when its module is built. The
elaboration of an import includes the evaluation of \lean|checkLrat|. Using
\lean|native_decide|, we obtain \lean|h| by evaluating
\lean|checkLrat cnfStr lratStr| as compiled code. The outcome is
asserted as one \emph{native-evaluation axiom}, named after the
declaration, which the kernel accepts without repeating the
evaluation and which subsequently appears among the axioms the
declaration depends on. Thus, the certificate is present as data and
does not contribute a proof structure. Rocq's native\_compute \citep{NativeCompute} also decides a proposition by running compiled code, where the compiler is trusted, and no axiom is asserted. lean-smt \citep{LeanSMT25} uses reflection, but its evaluation is checked by the kernel in terms of definitional equality. It does not invoke compiled code, and no axiom is asserted. \cv{Lean's three standard axioms are propositional extensionality,
choice, and quotient soundness.}%
\av{Continuing \Cref{ex:running}, we consider the theorem that states that the formula is unsatisfiable, which depends on Lean's three standard axioms (propositional extensionality, choice, and quotient soundness) and one native-evaluation axiom.}

With reflection, a single import evaluates \lean|checkLrat| once,
on the whole certificate. We refer to this as the \emph{whole-file
import}. The evaluation uses the checker's clause array, which grows
with the certificate size (compacted in \Cref{sec:compaction}), and
keeps the entire array in memory. The string \lean|lratStr| must
exist before the elaboration starts, and any interruption discards
the entire check. With large certificate sizes (\Cref{sec:eval}), the
whole-file import is infeasible due to memory limitations.

\subsection{Trust}\label{sec:trust}

A single import involves Lean, the SAT solver that produces the
certificate, and the mechanisms that prepare Lean's input. We refer
to this sequence of components as the \emph{import pipeline}. We
classify each component according to what a fault in it will cause.
A component is \emph{trusted} if a fault can lead to a false theorem,
and it is \emph{adversarial} if a fault can at most fail the build.
The basic trusted components are the Lean kernel with its three
standard axioms and the native-evaluation mechanism (compiler and
runtime). Our stream mode, which reads the certificate as streamed from a pipe
(\Cref{sec:streammode}), adds another trusted component, the byte
reader. It provides the pipe's data to the checker in blocks. The
adversarial components are the solver and its toolchain, with all
data they produce as files or streams. Other adversarial components are the snapshots
(\Cref{sec:runchunk}), which are serialized checker states and thus
data rather than code, and the renumberer (\Cref{sec:compaction}),
which rewrites certificate ids after the checker state has been
compacted. Recall that a fault in an adversarial component results
in the check evaluating to \lean|false| or in a failed elaboration.
We test this in \Cref{sec:evalT2}. Whereas the import pipeline starts with the formula, the
\emph{end-to-end pipeline} adds the encoding, which translates a
mathematical statement into the considered formula. The encoding is
often not included in the trust model, with the negative consequence
that an import could prove a true theorem about the wrong formula. We argue for including the encoding as an adversarial component within an end-to-end pipeline. Consequently, the
encoding should be written in Lean, with a Lean proof that
guarantees that unsatisfiability of the output implies the intended
mathematical statement. The import must refute exactly the encoder's
output. This is accomplished either by stating the theorem against
the term produced by the encoder, or by proving that the parsed
DIMACS file is contained in the list of clauses produced by the
encoder. A wrong encoding will then fail the
proof instead of misleading the reader. This way, the parser
\lean|parseDimacs| (\Cref{sec:reflection}) and the module generator
(\Cref{sec:chunked}) become adversarial. Consequently, only making
sure that the Lean statement matches the intended mathematical
statement remains to be trusted, which is the case for any theorem
in a proof assistant.
Our case studies (\Cref{sec:evalT3}) run the end-to-end pipeline.

Assume that the generator inserts an axiom or changes the statement
without stopping the build process. Then we would see the effect in
the theorem statement and in the printed list of axiom dependencies.

\section{An Overview of \tool{}}\label{sec:overview}     %

\begin{sloppypar}
The contributions of this paper are implemented in the reusable
tool \tool{}, which is a Lean~4 library and module generator that
imports any LRAT certificate independently of the underlying problem. The generator turns
the formula, its certificates, and the cube file into the Lean
modules of an import (\Cref{sec:chunked}).
\end{sloppypar}

To support an end-to-end pipeline, we state the problem in Lean, encode it into a formula $F$ within Lean, and prove the encoding's correctness in Lean as well (\Cref{sec:trust}).
The formula is then split into cases, or \emph{cubes}, with an
external cubing tool. Here, a cube is a conjunction of literals that are
fixed as assumptions for a single solver call. A \emph{leaf} is the formula that was cubed, with the cube's literals added as unit clauses. A
\emph{cover} is a list of cubes with the property that all the cubes
together provide a complete case distinction (\Cref{sec:cover}). The solver refutes each leaf and produces one certificate per leaf, plus one certificate that shows that the cubes form a cover. In practice, the cubing is often done after the solver has been run
on the input formula $F$ for a while. This run simplifies the
formula or adds symmetry-breaking clauses, and outputs the formula
$F'$. This derivation produces a certificate that guarantees that
unsatisfiability of $F'$ implies the unsatisfiability of $F$.

\tool{} imports each leaf independently, without reference to other leaves. The generator groups the leaves into modules, with a fixed number of leaves per module. Each module records the cube list of its leaves. Later, a top-level module checks that the cube lists of the modules concatenate to the list of the cover (\Cref{sec:cover}). This check is done by native evaluation. The checker (\Cref{sec:checker}) does not see the whole certificate at once. \tool{} divides the certificate of a single
leaf into chunks and checks each with respect to the state of the
previous chunk. Each chunk comes with a soundness lemma (Lemmas 2--4
of \Cref{tab:lemmas}), so that the chain of passed checks guarantees
that the leaf is unsatisfiable (\Cref{sec:checker}, the top of
\Cref{fig:chain}). The two import modes have the certificate located differently
during the check (\Cref{sec:streaming}). In file mode, the certificate is on disk, divided into chunks, each checked by its own chunk module. In stream mode, the certificate comes from the solver through a
pipe and is not stored. The stream mode checks the blocks in
sequence within one evaluation. It does not require a serialized
state. On the other hand, file mode restores the start state of each
chunk from a snapshot, and compaction rewrites the state through a
snapshot (\Cref{sec:compaction}). A top-level module uses Lemma~17 to combine the leaf theorems and
the theorem imported from the cover certificate, and states that $F$
is unsatisfiable (\Cref{sec:cover}, the bottom of \Cref{fig:chain}).
When we deal with a formula $F'$ derived from $F$ before the cubing,
\tool{} imports that derivation as a certified derivation and
processes it the same way, with Lemma~16 (\Cref{sec:deriv}), which
guarantees the unsatisfiability of $F$ based on the unsatisfiability
of $F'$.

Consequently, any \tool{} import, whether in file or stream mode,
runs in parallel and uses a bounded amount of memory. Both modes can deal with interruptions. In file mode, the chunks that already have been compiled into modules are kept. The import resumes with the rest. In stream mode, nothing is stored. So just the leaf that was interrupted is solved and checked again. When done, it lists all its axiom
dependencies, including the axioms emerging from the import. We can build the modules in parallel since each
module depends only on the module that defines $F$. The largest
amount of memory required in such a process is that of one leaf.
The final theorem depends on Lean's three standard axioms and one
native-evaluation axiom per evaluation. That includes one axiom per
chunk in file mode or one axiom per leaf in stream mode, one for the
cover, and one axiom for the check that the modules' cube lists
concatenate to the cover's list (\Cref{sec:cover}).

\begin{figure}
\centering
\begin{tikzpicture}[>=stealth, x=1cm,
  chunk/.style={draw, rounded corners=1pt, inner sep=3pt, font=\small},
  leaf/.style={draw, rounded corners=1pt, inner sep=3pt, font=\small,
    minimum width=1.1cm},
  st/.style={font=\small\ttfamily\color{leancode}},
  lem/.style={font=\small\ttfamily\color{leancode}},
  lab/.style={font=\small}]
\node (F)  at (0,0)   {$L_2$};
\node[chunk] (c1) at (1.25,0) {$c_1$};
\node (s1) at (2.5,0) {$s_1$};
\node[chunk] (c2) at (3.75,0) {$c_2$};
\node (s2) at (5.0,0) {$s_2$};
\node[chunk] (c3) at (6.25,0) {$c_3$};
\node (bot) at (7.4,0) {$\bot$};
\draw[->] (F) -- (c1);  \draw[->] (c1) -- (s1);
\draw[->] (s1) -- (c2); \draw[->] (c2) -- (s2);
\draw[->] (s2) -- (c3); \draw[->] (c3) -- (bot);
\node[st, above=3pt of c1] (t1) {stepStart};
\node[st, above=3pt of c2] {stepMid};
\node[st, above=3pt of c3] (t3) {stepFinish};
\draw[->, dashed] (s1.south) to[bend left=40]
  node[lab, below] (b1) {Lemma 2} (F.south);
\draw[->, dashed] (s2.south) to[bend left=40]
  node[lab, below] {Lemma 3} (s1.south);
\draw[->, dashed] (bot.south) to[bend left=40]
  node[lab, below] (b3) {Lemma 4} (s2.south);
\node[draw, dotted, rounded corners=2pt, inner sep=4pt,
  fit=(F) (bot) (t1) (t3) (b1) (b3)] (frame) {};
\node[lab, anchor=south west, inner sep=1pt] at (frame.north west)
  {import of leaf $L_2$: a chain of chunk checks};
\node[leaf] (l1) at (0.6,-2.6) {$L_1$};
\node[leaf] (l2) at (2.1,-2.6) {$L_2$};
\node[lab]  (ld) at (3.3,-2.6) {$\cdots$};
\node[leaf] (lm) at (4.5,-2.6) {$L_m$};
\node[leaf] (cov) at (6.4,-2.6) {cover};
\node[st] (base) at (3.7,-3.8) {base.Unsat};
\draw[->, dashed] (l1.south) -- (base.north west);
\draw[->, dashed] (l2.south) -- (base.north);
\draw[->, dashed] (lm.south) -- (base.north);
\draw[->, dashed] (cov.south) -- (base.north east);
\node[lab, right=2pt of base] {Lemma 17};
\draw[dotted] (frame.south) -- (l2.north);
\node[lab, anchor=north, inner sep=1pt] at (base.south)
  {the leaf theorems and the cover theorem compose};
\end{tikzpicture}
\caption{The two levels of an import. Top: one leaf's certificate
is checked as a chain of chunk checks (solid) whose step lemmas (dashed, in the direction of the implication) compose into the leaf's unsatisfiability;
$s_1$ and $s_2$ are the boundary snapshots, $\bot$ the empty
clause. Bottom: the leaf theorems and the cover theorem compose by
Lemma~17 into the theorem about $F$. In file mode each chunk is a
Lean module, in stream mode each group of leaves; Lean's build tool Lake builds the modules in parallel.}
\label{fig:chain}
\end{figure}
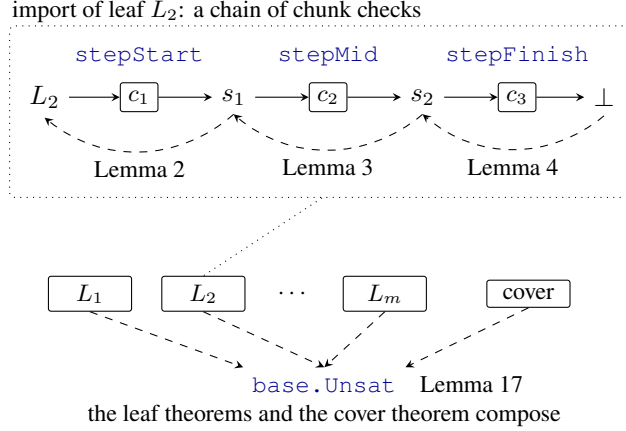
\begin{table*}
\caption{Import mechanisms for computational results; the first two mechanisms are Isabelle's and Rocq's (formerly Coq), the others Lean's. ``std'' =
Lean's three standard axioms; a dash means the notion does not
apply.}
\label{tab:modes}
\scriptsize
\setlength{\tabcolsep}{\modesSep}
\renewcommand{\arraystretch}{1.05}
\newcolumntype{C}{>{\raggedright\arraybackslash\tightlines}p{\modesCol}}
\begin{tabular}{@{}>{\raggedright\arraybackslash\tightlines}p{\modesFirst}CCCCCCC@{}}
\toprule
& \emph{oracle} & \emph{native\_\hspace{0pt}compute} & \emph{cake\_lpr pipe}
  & \emph{re-solve at elaboration} & \emph{whole-file import}
  & \textbf{file mode} & \textbf{stream mode}\\

\midrule
prover receives & theorem (oracle-tagged) & theorem
  & axiom & theorem & theorem & theorem & theorem\\
general tool or one project? & general & general & one project & both (see text) & general & general & general\\
rebuildable from kept data? & no (rerun tool) & yes (re-eval.)
  & no (re-solve) & no (re-solve) & yes & yes & no (re-solve)\\
certificate recheckable later? & --- & --- & no & no & yes
  & embedded copy & no\\

trusted code added & whole tool & compiler chain
  & axiom-asserting code & native eval. & native eval. & native eval.
  & native eval.\ + byte reader\\

\midrule
axioms of the theorem & none (oracle tag instead) & none
    (trust implicit) & std + one axiom per verdict & std + one native axiom
  & std + one native axiom
  & std + native, one per chunk & std + native, one per run\\
\bottomrule
\end{tabular}
\end{table*}

\subsection{Comparing import mechanisms}\label{sec:modes}

In \Cref{tab:modes}, we compare our two import modes and the
whole-file import (\Cref{sec:reflection}) with four published
mechanisms. The row on trusted code lists the code that is added to the prover's trusted base if an unverified external tool
produces a certificate and a verified checker inside the prover
validates it.

An Isabelle \emph{oracle}, as \citet{Weber09} describe, admits an external tool's answer as an oracle-tagged theorem, and a Rocq \emph{native\_compute}
evaluation \citep{NativeCompute} decides a proposition by running
compiled code; both can produce a false theorem if the code they depend
on is faulty. The
\emph{cake\_lpr pipe} is the pipeline with which
\citet{Subercaseaux24} formalized the computation of the empty-hexagon number of \citet{HeuleScheucher24}. They re-ran the solvers,
piped each certificate to cake\_lpr as it was produced, stored
nothing, and recorded the verdict in Lean as an axiom. \citet{Gocht24} check VeriPB's pseudo-Boolean certificates with their verified checker CakePB. In this setup, no theorem prover is involved. Their elaborator writes proof files to disk. However, they also verified the encoder for subgraph solving, so their toolchain is end-to-end. CakePB's verdict could be recorded in a prover similar to the cake\_lpr pipe. The symmetry and dominance breaking that is certified
in VeriPB by \citet{Bogaerts23} can be considered a counterpart of
our certified derivations (\Cref{sec:deriv}). \emph{Re-solve at
elaboration} means the solver is run at elaboration time and its
certificate is checked by reflection. Lean's \lean|bv_decide| tactic
\citep{Boving25} does this as a general tool, using the verified
checker of Lean core. \citet{Codel26} did this when they formalized the Keller conjecture, with a project tactic in their published artifact for $3 \le d \le 5$.
The remaining columns, whole-file import, file mode, and stream
mode, also check the certificate inside the prover at evaluation
time; however, the solver runs outside of it. With whole-file
import, the stored certificate is loaded as a whole
(\Cref{sec:reflection}). Our file mode imports a stored certificate
in chunks and keeps each chunk's text in its module
(\Cref{sec:chunked}). Our stream mode reads the certificate from a
pipe to which the solver writes it; nothing is stored
(\Cref{sec:streammode}). We are not aware of another system that translates certificates that are too large for a whole-file import (\Cref{sec:reflection}) into a theorem.

A native-evaluation axiom asserts that one closed Boolean term
evaluated to \lean|true| under Lean's compiled code. Lean adds one
such axiom per evaluation. Consequently, the axioms a composed
theorem depends on list all the evaluations, which are hundreds of
thousands in the largest imports we consider (\Cref{sec:evalT3}).
All these axioms are of the same form and use the same trusted base,
which is the compiler and the runtime, on which \lean|bv_decide|
relies as well.
\av{Each of these axioms coming from individual evaluations is not a separate trust assumption. }%
The trusted code is the same whether it is one
such axiom or a million. One could group leaves into one evaluation
to reduce many of these axioms to a single one, but we decided against this cosmetic improvement.
\av{A verdict axiom of the cake\_lpr pipe is as trustworthy as the code that turns the verdict into an axiom, which is listed in the table row ``trusted code added''.}

In contrast to cake\_lpr, which is verified down to the machine
code the CakeML compiler produces, Lean's compiler and runtime are
not verified. Depending on the import, we change what is trusted,
not the size of the trusted base.
\av{The native-evaluation axiom asserts one specific fact: that a closed Boolean term evaluated to \lean|true|. }%
The closed term of a native-evaluation axiom captures the run of a
checker verified in Lean. Hence, this is the same system in which
the theorem is formulated. The formula the checker was applied to is
exactly the one named in the theorem statement. By contrast, the verdict axiom of the cake\_lpr pipe
asserts whatever proposition the code that writes the axiom states.
Importantly, the link between the checker's verdict and that
proposition is in that code, not in a proof.

To check a certificate while the solver runs is an established
technique outside the context of proof assistants. \citet{Lammich24}
verified lrat\_isa in Isabelle down to LLVM~IR. lrat\_isa is run
alongside the solver, and it checks the LRUP fragment of LRAT (RUP
additions only). \citet{Schreiber24} ran an unverified companion checker on the native LRAT output that CaDiCaL writes
\citep{Pollitt23, CaDiCaL24}. \citet{Battleman25} partition the problem using splitting variables from a proof prefix. \citet{Michaelson25} build one certificate from the traces of clause-sharing solvers. Our approach divides the checking of one certificate.

\section{A Resumable Verified Checker}\label{sec:checker}         %

\av{\enlargethispage*{3mm}}%
A whole-file import (\Cref{sec:reflection}) evaluates
\lean|checkLrat| once, on the full certificate. In this section we divide that evaluation into one check per segment of the certificate. The segments are linked only by serialized states of
the checker, and nothing validates those before use. Each check's
statement mentions only the states it starts from and ends in. The
soundness lemmas hold for any state, so a wrong serialization can at
most fail the build.

\subsection{Per-chunk checks and their composition}\label{sec:chaining}

A \emph{chunk} $c$ is a contiguous segment of the certificate under
consideration. It contains consecutive actions in certificate
order, parsed into the checker's action type \lean|IntAction|. An
import with chunks $c_1, \ldots, c_k$ amounts to $k$ checks, each of
which maps its chunk to a Boolean value. The computation is divided
at \emph{boundaries}. At a boundary, the checker's state is given to
the next check as a \emph{snapshot}, consisting of a serialized copy
of its clause array (\Cref{sec:runchunk}). It depends on the import mode (\Cref{sec:streaming}) whether each
check is a single native evaluation or all checks together are one.
The first chunk starts from $F$ (\lean|stepStart|, which builds its
initial state from $F$ instead of a snapshot) and must end in its
output snapshot $s_1$. A middle chunk resumes from its input
snapshot $s$ and must end in its declared output snapshot $s'$
(\lean|stepMid|\cv{, \Cref{app:displays}}). The last chunk must derive the empty clause
(\lean|stepFinish|, which requires the result \lean|refuted|
instead of comparing snapshots). \lean|restoreD| and \lean|serialize| convert between snapshots and
checker states, and \lean|runChunk| runs the chunk on a state
(\Cref{sec:runchunk}).
The middle check \lean|stepMid|, which has
\lean|n| as the variable bound:

\begin{lstlisting}
def stepMid (n : Nat) (prev : Snapshot)
    (chunk : Array IntAction)
    (next : Snapshot) : Bool :=
  match runChunk (restoreD n prev) chunk 0 with
  | some (.more f') => serialize f' == next
  | _ => false
\end{lstlisting}

For each check, there is a soundness lemma that we refer to as a
\emph{step lemma}; see \Cref{fig:chain}. If a middle check passes, it proves that
unsatisfiability of the state restored from $s'$ implies the
unsatisfiability of the state restored from $s$ (Lemma~3\cv{, \Cref{app:displays}}), and a
passed last check makes the state restored from $s$ unsatisfiable (Lemma~4\cv{, \Cref{app:displays}}). If the first check passes, it proves that the
unsatisfiability of the state restored from $s_1$ implies the
unsatisfiability of $F$ (Lemma~2). Composed from
the last chunk back to the first, the step lemmas prove that $F$ is unsatisfiable by a proof term that uses no further native evaluation.
The statements of Lemmas 3 and 4; the
snapshots $s$ and $s'$ of the text appear as \lean|prev| and
\lean|next|:

\begin{lstlisting}
theorem stepMid_sound  -- Lemma 3
    (h : stepMid n prev chunk next = true) :
    Unsatisfiable (PosFin n) (restoreD n next) →
    Unsatisfiable (PosFin n) (restoreD n prev)

theorem stepFinish_sound  -- Lemma 4
    (h : stepFinish n prev chunk = true) :
    Unsatisfiable (PosFin n) (restoreD n prev)
\end{lstlisting}

A wrong snapshot fails the build at the first check that uses it. In case it still checks, the composed statement can
only mention $F$ itself and is true. Our fault injections
(\Cref{sec:evalT2}) confirm this.

\subsection{Resuming from any snapshot}\label{sec:runchunk}

A chunk may start from a snapshot that nothing has validated
(\Cref{sec:chaining}). A \emph{checker state} is the core checker's
state (type \lean|DefaultFormula n|, with every variable below
$n$): a clause array indexed by clause id, plus two unit-propagation
caches and an assignment array. The checker keeps the assignment array consistent with the clause array. A snapshot is the clause array
alone, as a list. Since hints are indexed by id, a deleted clause
leaves an empty entry, a \emph{hole}. \lean|serialize|
produces the snapshot, and \lean|restoreD| rebuilds a state from it
and the bound $n$\cv{ (snapshot module in \Cref{app:displays})}. The three checks call the function
\lean|runChunk|, which processes the actions of a chunk on a state
the same way the core checker does. However, it returns one of three
results: the final state \lean|more|, the verdict \lean|refuted|
that the chunk derived the empty clause, or the failure \lean|none|.
The snapshot module at the first boundary of \Cref{ex:running}, in
its textual form (a dot is a hole; clauses render in internal
literal order):

\begin{lstlisting}
def snap1 : LRATCatcher.Snapshot :=
  LRATCatcher.parseSnap ".
1 2 3 0
.
-2 1 0
2 -1 0
-1 -2 3 0
-1 -2 -3 0
2 1 0
1 0
"
\end{lstlisting}

\begin{sloppypar}
Unsatisfiability depends only on the clause array. Consequently, a
state and its restored serialization agree on it (Lemma~7). The soundness lemmas of the core checker require a \emph{ready} state, that is, a state satisfying \lean|ReadyForRupAdd| and \lean|ReadyForRatAdd|. This means that the two unit-propagation caches are empty. Every recorded assignment corresponds to a unit clause that is present in the clause array. Since \lean|restoreD| builds the state with the core's constructor
\lean|ofArray|, which the core proves ready for arbitrary clause
arrays, every restored snapshot is ready (Lemma~8), whether or not
it is the serialization of an actual state. A chunk that derives the
empty clause from a ready state establishes that the state is
unsatisfiable (Lemma~5). A chunk that completes without deriving the
empty clause returns a ready state whose unsatisfiability implies
the unsatisfiability of the input (Lemma~6\cv{, \Cref{app:displays}}). The proof of
Lemma~6 proceeds by induction over \lean|runChunk|'s loop. Each
case is proved by a per-step lemma of the core checker. A RUP
addition yields a model-equivalent state (\lean|rupAdd_sound|), a
RAT addition an equisatisfiable state (\lean|ratAdd_sound|), and a
deletion a state that is implied by the original state
(\lean|limplies_delete|). The step lemmas follow directly from
Lemmas 5--8.
\end{sloppypar}
The statement of Lemma~6:

\begin{lstlisting}
theorem runChunk_more_sound  -- Lemma 6
    (h1 : Formula.ReadyForRupAdd f)
    (h2 : Formula.ReadyForRatAdd f) :
    runChunk f chunk idx = some (.more f') →
      Formula.ReadyForRupAdd f' ∧
      Formula.ReadyForRatAdd f' ∧
      (Unsatisfiable (PosFin n) f' →
       Unsatisfiable (PosFin n) f)
\end{lstlisting}

\section{Import Modes: File and Stream}\label{sec:streaming} %

The chunk checks and boundaries as discussed above support two ways
of turning a certificate into a theorem. They differ in where the
certificate bytes are located while they are checked. In file mode,
they are located on disk, divided into chunks; in stream mode, they
come through a pipe and are not stored.

\subsection{File mode}\label{sec:chunked}

The \emph{module generator} turns an input formula $F$ and the %
certificate into Lean modules. Each module is a compiled file that
contains its statement and, for a chunk, the chunk's textual
content. We have four kinds of modules. The formula module contains
the parsed formula, called \lean|base| in the generated code. A
snapshot module exists for each boundary, and a chunk module for
each chunk. The closing module composes the step lemmas into the
theorem \lean|base_unsat : base.Unsat|. Since the formula module is the generated Lean file that contains
the CNF term, it determines which formula the theorem is about. In
the end-to-end pipeline, it is adversarial since its term must be
equal to the encoder's term or be proved contained in it
(\Cref{sec:trust}). The \emph{chunk size} $L$ is the number of
certificate actions that one module covers. The seven actions of \Cref{ex:running} form chunks of three,
three, and one action ($L = 3$). The module of the second chunk states \lean|stepMid nv snap1 (parseActions "...") snap2 = true| and proves it by \lean|native_decide|\cv{ (\Cref{app:displays})}. Since each module stores its chunk's actions as a string literal,
the original certificate is not needed once the modules exist. A
chunk module imports only the formula module and the two snapshot
modules it involves, not the preceding chunk module. Consequently, the chunk modules are independent. After an interruption, only the modules
without a compiled artifact need to be rebuilt, each by resuming
from its stored input snapshot. Hence, $L$ defines how much
work a resume needs to repeat (\citet{Cirpons26} report a batch-size trade-off for reflection). For this import, the only
axiom dependencies of \lean|base_unsat| are Lean's three standard
axioms and one native-evaluation axiom per chunk, \lean|step1| to \lean|step3|\cv{ (\Cref{app:displays})}.
The module of the second chunk of \Cref{ex:running}:

\begin{lstlisting}
set_option maxHeartbeats 0 in
theorem step2 : LRATCatcher.stepMid nv snap1
  (LRATCatcher.parseActions "9 -1 -2 0 5 6 0
9 d 7 0
10 -1 0 9 4 0
") snap2 = true := by native_decide
\end{lstlisting}

\noindent
The option removes Lean's elaboration step limit. \lean|nv| is the
variable bound $n$ of this formula, generated with the formula module,
\lean|parseActions| the certificate parser applied to the chunk's
text, and \lean|snap1| and \lean|snap2| the two boundary snapshots.

The complete list of axiom dependencies of \lean|base_unsat| for this
import, module prefixes omitted:

\begin{lstlisting}
'base_unsat' depends on axioms:
[propext, Classical.choice, Quot.sound,
 step1._native.native_decide.ax_1_1,
 step2._native.native_decide.ax_1_1,
 step3._native.native_decide.ax_1_1]
\end{lstlisting}

\subsection{Stream mode}\label{sec:streammode}

In stream mode, the certificate is checked while it is streamed
from the solver. No data is written to persistent storage. A
\emph{feed} is a function from a block index to the actions of that
block (or to nothing once the stream has ended). The loop
\lean|checkStream| reads blocks from the feed and runs each with
\lean|runChunk| (\Cref{sec:runchunk}) on its current state until
one block refutes. The current state is \lean|f| at entry in
Lemma~9 below, and \lean|i| is the block index, zero at the start.
A \emph{feed block} is a run of at most $B$ consecutive certificate
actions, written one per line in the textual format. $B$ is fixed by
the code that reads the pipe (described below). Since Lemma~9
quantifies over the feed, it holds for any chosen $B$. Since Lean accepts a recursive function only if it is evident that
the recursion terminates, we arrange this for \lean|checkStream|
with an extra argument \lean|fuel|, a natural number. Every
recursive call passes \lean|fuel - 1|, so the recursion is on a value that strictly decreases.
\av{Consequently, Lean accepts the definition without any termination argument about the stream. }%
If the counter is exhausted, if the stream ends without producing a refutation, or if a check fails, the loop returns \lean|false|:

\begin{lstlisting}
theorem checkStream_sound  -- Lemma 9
    (feed : Nat → Option (Array IntAction))
    (fuel : Nat) :
    ∀ (f : DefaultFormula n) (i : Nat),
      Formula.ReadyForRupAdd f →
      Formula.ReadyForRatAdd f →
      checkStream feed f fuel i = true →
      Unsatisfiable (PosFin n) f
\end{lstlisting}

\noindent
The import fixes \lean|f| to the initial state built from $F$,
similar to \lean|stepStart|. A successful run elaborates the theorem stating
that the CNF term of $F$ is \lean|.Unsat| (\Cref{sec:reflection}).
\av{The theorem depends on Lean's three standard axioms and one native-evaluation axiom for the run. }%
Lemma~9 quantifies over \lean|feed| and consequently holds for
every feed, however garbled or adversarial. The actual stream does not need
to be trusted. At checking time, \lean|feed| is instantiated with an
\lean|opaque| constant. Its implementation composes a small
\emph{byte reader}, which carries out buffered reads from the pipe
using Lean's IO primitives, with the certificate parser, also
written in Lean. We require opaqueness for the following reason. If
the feed were an ordinary definition with a compiled implementation
that differs from that definition, then native evaluation would
assert a proposition about the definition, and that proposition
could be provably false.
\av{Note that the implementation reads the pipe; the definition does not. }%
However, an opaque constant has no
definition, so there is nothing to be contradicted. The
native-evaluation axiom records the evaluated proposition, and that
proposition is satisfied by some feed. Some feed satisfying it does not alone make a theorem true; it does here because the conclusion of Lemma~9 does not mention the feed. A run started at block $i$ with
counter value $\mathit{fuel}$ reads the feed only at the indices in
$[i, i + \mathit{fuel})$, and Lemma~10 proves that two feeds that
agree on those indices give the same result. Consequently, if a run returns true, it will also return true on every feed that extends the blocks it has already read. The native-evaluation axiom records that the run returned true. Lemma~9, if applied to any such feed, turns this into the statement that $F$ is unsatisfiable. The only step outside
Lean is that the implementation of the opaque constant returned
those blocks. Each import declares its own feed constant; all share one audited implementation.
\av{Hence, a rebuild replaces the theorem, its feed constant, and its axiom together in the module's compiled artifact.}

The only trusted component that is added by the stream mode is the
byte reader. Whatever it returns, faulty or scrambled or even missing, mirrors the behavior of some feed. Consequently, what remains trusted is only that its 171 lines of Lean code do not corrupt the evaluation.
\av{We add the block index as an argument to the byte reader. This ensures that every read gets a different call, and the Lean compiler does not evaluate a closed term once and reuse the result.}

There is no completeness or termination theorem, on purpose. The
counter bounds the loop, and no liveness is assumed of the feed.

\subsection{Boundary compaction}\label{sec:compaction}

The only place in the checker that grows with certificate size is
the clause array. With additions and deletions, it is modified
constantly. Ids are positions in it, so the array is as long as the
largest id ever allocated. As an example, on the 276~MB and 491~GB
certificates we analyze in \Cref{sec:eval}, the largest numbers of
live clauses (added and not yet deleted) at any time were 2.1\% and
0.20\% of all clauses added, respectively. All other allocations of
the checker's persistent state are bounded by the formula and the
live set. The buffers have fixed sizes. This bound is not
theoretically shown; we test it per instance in \Cref{sec:evalK}.
With \emph{boundary compaction}, we remove the growing term by
dropping the holes at a boundary (\lean|compactSnapshot|). In stream
mode, the state is serialized and immediately restored at such a
boundary, within one evaluation. The compaction lemma, Lemma~12\cv{ (\Cref{app:displays})}, states that restoring a compacted snapshot yields a state that is model-equivalent (\lean|Liff|) to the state it was taken from. Note that dropping the holes preserves the list of live clauses. The lemma has two hypotheses, one per propagation
cache, that require the cache to be empty. This holds for every
state that is returned by a successful chunk run (Lemmas 6 and 11). There is a compacting variant of the stream loop,
\lean|checkStreamC|, which interleaves this step every $K$ feed
blocks. We call $K$ the \emph{compaction interval}. Its soundness
lemma, Lemma~13, applies Lemma~12 once per boundary. Lemma~13 takes $K$ as a parameter. By choosing $K$, we affect only speed and memory, not
soundness.
\Needspace*{12\baselineskip}
The compaction round-trip lemma, Lemma~12:

\begin{lstlisting}
theorem liff_restoreD_compactSnapshot_serialize
    -- Lemma 12
    (f : DefaultFormula n)
    (hrup : f.rupUnits = #[])
    (hrat : f.ratUnits = #[]) :
    Liff (PosFin n)
      (restoreD n (compactSnapshot (serialize f)))
      f
\end{lstlisting}

Compaction changes the ids. After a boundary, the live clauses get
new positions. Hence, the ids and hints of certificate actions that
appear later must be renumbered to match. Therefore, we use a
\emph{renumberer}, a streaming preprocessor which rewrites the ids
and hints. It recomputes the checker's array positions from the same
certificate, so it does not need information from the Lean process.
The renumberer is an adversarial component (\Cref{sec:trust}), since it rewrites certificate ids but not the formula. \cv{\Cref{app:displays}
shows the renumbered continuation of the certificate of
\Cref{ex:running}. }All file-mode imports that we consider in this
paper are uncompacted.
The renumbered continuation of the certificate of \Cref{ex:running}
under compaction after every third action ($B$ = one action,
$K = 3$):

\begin{lstlisting}
8 -1 -2 0 4 5 0    -- was id 9; hints 5 6
9 d 6 0            -- was 9 d 7 0 (label ignored)
9 -1 0 8 3 0       -- was id 10; hints 9 4
9 0 6 8 0          -- was id 11; hints 8 10
\end{lstlisting}

\noindent
After the first compaction the id sequence restarts at the compacted
position count (here: 8 again, immediately after ids had reached 8);
the number before \lean|d| on a deletion line is a label the
checker ignores; and the final line's id 9 is a reuse, because a
second compaction occurs after the third line, while its hints are remapped twice, once per compaction.

\section{Composition: Cube-and-Conquer and Derivations}\label{sec:compose} %

For large instances, a refutation of a formula $F$ consists of many
certificates, one per cube and one showing that the cubes cover all
assignments. In this section, we compose import theorems across the
leaves of a cube-and-conquer cover and across a preprocessing stage.
The certificate of a preprocessing stage certifies the
transformation from $F$ to $F'$ instead of refuting $F$.

\subsection{Composing leaves}\label{sec:cover}

Lemma~17, displayed below, composes the theorems of the leaf
certificates and the cover certificate into a proof of the
unsatisfiability of $F$. Here, a cube is a conjunction of literals
that are fixed as assumptions. A leaf is the formula obtained by
adding the unit clauses of a cube to $F$. We also say leaf for the leaf's certificate and its import. A cover is a list of cubes
with the property that every assignment satisfies some cube. The hypothesis \lean|hcover| states that the cubes indeed
form a cover. Negating the literals of a cube gives us a clause. The
conjunction of the clauses of all cubes gives us a formula, which is
unsatisfiable exactly when every assignment satisfies some cube, the
cover property. A refutation certificate of this formula is
therefore a certificate for the cover.

\begin{lstlisting}
theorem cover_unsat  -- Lemma 17
    (hleaf : ∀ c ∈ cubes,
      (Cube.leafCNF c base).Unsat)
    (hcover : (negCubesCNF cubes).Unsat) :
    base.Unsat
\end{lstlisting}

\noindent
Here \lean|base| is a bound variable that gets instantiated at $F$,
not the constant \lean|base| of \Cref{sec:chunked};
\lean|Cube.leafCNF c base| is the leaf of cube \lean|c|, and
\lean|negCubesCNF cubes| denotes the cover formula. Each leaf gives rise to a separate import and theorem, stated by
the generator as \lean|(Cube.leafCNF c base).Unsat|. With Lemma~17,
we compose the theorems. In \Cref{sec:checker}, we divided the
checking of one certificate into chunks and joined the chunk
theorems with step lemmas. Lemma~17 accomplishes the same one level
up: it joins the theorems of separate certificates. This is
possible since each leaf yields a theorem; a checker's verdict is a Boolean and cannot be composed. The leaves themselves can be streamed, chunked, or compacted
imports. Each of them ends in \lean|.Unsat| of a
statement-level \lean|CNF Nat| (\Cref{sec:reflection}), which,
instantiated at \lean|Cube.leafCNF c base|, is a leaf hypothesis
(\lean|hleaf|). The cube list is parsed inside Lean from the cube file. This parser is adversarial, since Lemma~17 holds for any list
of cubes; so a wrong parse can only fail the cover check or a leaf
check. The
cover certificate is embedded in its module. We check it as a
whole-file import (\Cref{sec:reflection}). The generator groups the
leaves into modules of a fixed number of leaves. It then proves by
native evaluation that the groups' cube lists concatenate to the
parsed list, which yields one more native-evaluation axiom.

\subsection{Certified derivations}\label{sec:deriv}

The certificates considered so far all ended in the empty clause.
An import whose chunks (\Cref{sec:chaining}) end without the empty
clause is a \emph{certified derivation}. Its additions and deletions
transform $F$ into a derived formula $F'$. Such an $F'$ is typically
produced by a solver's preprocessing. The last chunk of such a
certificate is checked by \lean|stepMid| and ends in a final
snapshot. The next stage splits and refutes $F'$ outside Lean, and
no copy of $F'$ outside Lean is trusted. The $F'$ that the later
theorems refer to, however, is computed inside Lean:
\lean|externSnap| computes $F'$ from the final snapshot; thereby it
drops the holes and renders the clauses in the input's variable
numbering as a statement-level \lean|CNF Nat|. Every later statement mentions exactly that term. The DIMACS file
given to the solver is a rendering of it. Any mismatch can fail the
build but will not yield a theorem about a wrong formula, as in
\Cref{sec:chaining}. The refutations of the leaves are imported as
a cover (\Cref{sec:cover}), and elaboration composes the theorems.

Lemma~15\cv{ (\Cref{app:displays})} relates an external refutation to the chain of step
lemmas. Its hypothesis is that the formula \lean|externSnap snap|,
which consists of the snapshot's clauses considered as a plain
formula, is unsatisfiable. This is what the cover of the next stage proves. The lemma's conclusion is that the restored checker state
\lean|restoreD n snap| is unsatisfiable, which is what the step
lemmas of \Cref{sec:chaining} take as input. Lemma~15 also needs a
side condition, that the snapshot decodes successfully, so that the restored
state and the plain formula hold the same clauses. Every snapshot
that a successful step check produced satisfies this condition
(Lemma~14). We need this side condition because \lean|restoreD| returns the
empty, satisfiable formula when decoding fails.
\av{For the empty formula, all the step lemmas hold vacuously, while \lean|externSnap| of the same snapshot might be unsatisfiable. }%
The purpose of
Lemma~16 is to compose the implication with the step lemmas
(\Cref{sec:chaining}), to derive the unsatisfiability of $F$ from
the external refutation.
The statement of Lemma~15:

\begin{lstlisting}
theorem extern_unsat_restoreD  -- Lemma 15
    (hres : (restoreList n snap).isSome)
    (hx : (externSnap snap).Unsat) :
    Unsatisfiable (PosFin n)
      (restoreD n snap)
\end{lstlisting}

Our adversarial solver stages write entire formulas from one
process to the next, having each stage evidenced by its own
certificate and composed by theorem dependency. This differentiates
our approach from the one by \citet{Codel26}, who validate
substitution-redundancy additions with a verified checker and
transform one formula in place in their symmetry-breaking stage.

\section{Evaluation and Case Studies}\label{sec:eval}             %

All experiments ran on a machine with 48 cores and 811~GB of memory,
except for the cluster runs noted in \Cref{sec:evalHex,sec:evalT3}.
We used Lean v4.30.0 and CaDiCaL~3.0.0, where CaDiCaL is run with \lean|--no-factor| so that fresh variables are not introduced, as we support RAT steps only on the input variables.
We report memory in terms of the largest amount of main memory the
Lean process occupied at any point of the run (RSS in the tables). About 1.5~GB of this
is the process itself, before any certificate is processed. The rows
of the tables are single runs unless a caption says otherwise. \cv{We provide in the supplementary archive the \tool{} code, and the
scripts and solver invocations that regenerate the cube files and
Lean modules of the case studies. It also contains the composed
modules of the small case studies, the axiom dependencies, and a map
from every lemma of \Cref{tab:lemmas} and every theorem of
\Cref{tab:cases} to its source file. The generated modules of the
large case studies (12~GB of Lean sources) and the retained
certificates (385~GB) are too large and therefore not included.
However, we provide scripts that produce these modules and
certificates. All code and data will be made public upon
publication.}%
\av{The tool is available in the public repository
\url{https://github.com/leansolving/lrat-catcher}. The repository
also contains a small end-to-end example, the Schur number
$S(4) = 44$. It runs on a single core in a few minutes and includes
the Lean encoding, split into eight cubes, solving, and streamed
import to the composed theorem. The supplementary material of the
case studies is available on Zenodo (\url{https://doi.org/10.5281/zenodo.22341464}). It contains scripts
and solver invocations to reproduce the case studies.}

\subsection{The same certificates across tools and modes}\label{sec:evalT1}

To import a certificate as a theorem costs about 15 times the CPU
time of the fastest verified verdict, plus one Lean process.
However, the import produces a declaration with stated axiom
dependencies, which is not possible with a verdict. In
\Cref{tab:tools}, we compare stream mode, the whole-file import (\Cref{sec:reflection}), Mathlib's \lean|lrat_proof|, and three external checkers on three
certificates, ranging in size from 63~MB to 58.7~GB. These are the
PHP(10,9) pigeonhole formula and two bounds $R(B_s,B_t) \le n$ on
Ramsey numbers of book graphs \citep{Wesley26}, where the CNF at
order $n$ is unsatisfiable exactly when the bound holds. We use lrat\_isa
(\Cref{sec:modes}), the fastest verified checker, as the reference
for the factor of 15 mentioned above; it reads the binary form of
the 276~MB certificate, as does our stream mode in that comparison.
cake\_lpr (\Cref{sec:certificates}) is also verified. lrat-check is
the unverified reference checker.

\begin{figure}
\centering
\av{\includegraphics{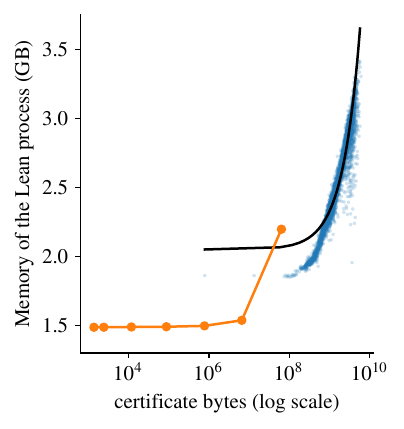}}
\caption{Memory of the Lean process against certificate bytes (log scale).
Blue: the 4,882 group imports of the empty-hexagon re-solve
(\Cref{sec:evalHex}), each group's memory against the largest
single leaf certificate it contains; Pearson $r = 0.94$, with the
linear fit (black line) at 2.05~GB plus 0.28~GB per GB of largest
leaf. Orange: whole-file imports of the PHP$(n,n-1)$ instances,
flat at the process's own size until the certificate itself grows. The
fit and $r$ are computed on untransformed bytes; the intercept
exceeds the process's own size by the formula stored in the checker,
about 0.4~GB for the hexagon CNF's 436,523 clauses.}
\label{fig:scatter}
\end{figure}

Comparing the whole-file import with stream mode on the same certificate, we find that the former needs about three times the memory and
twice the CPU time of stream mode. This is caused by it keeping the
certificate string and the parsed actions in memory all at once.
With the division of \Cref{sec:checker}, we avoid this requirement.
The memory requirement of a streamed import grows with the number of actions that are present in the certificate. It reaches 4.87~GB on the 58.7~GB certificate. With boundary
compaction, we can avoid this growth (\Cref{sec:evalK}).
\Cref{fig:scatter} shows the two parts of this memory requirement:
the process size with the live clause set, and the growth with the
size of the largest leaf.

\begin{table}
\caption{Certificate checking across tools and modes on one machine,
ASCII certificates unless noted. Our rows are stream mode unless labeled; the
276~MB rows of ours are medians of five or six runs; lrat\_isa reads
binary LRAT, and the ``stream (bin.)'' row reads the same binary
form; cake\_lpr's memory on the 58.7~GB certificate is the smallest
heap setting that verified it (a 4.3~GB heap ran out).
``Produces'' distinguishes a verdict (an exit code) from a theorem (\Cref{tab:modes}).}
\label{tab:tools}
\footnotesize
\setlength{\tabcolsep}{3pt}
\renewcommand{\arraystretch}{1.05}
\begin{tabular}{@{}lrrrl@{}}
\toprule
tool or mode & wall & CPU (s) & RSS (GB) & produces\\
\midrule
\multicolumn{5}{@{}l@{}}{PHP(10,9): 63.4 MB; wall in s}\\
stream                & 15.8 & 3.24 & 1.52 & theorem\\
lrat\_isa (bin.)      & 0.16 & 0.14 & 0.01 & verdict\\
cake\_lpr             & 3.23 & 1.95 & 2.85 & verdict\\
lrat-check            & 0.83 & 0.80 & 0.04 & verdict\\
\lean|lrat_proof|     & ---  & ---  & out of memory & none\\
\addlinespace
\multicolumn{5}{@{}l@{}}{$R(B_2,B_9) \le 22$: 275.9 MB; wall in s}\\
stream                & 20.9 & 13.1 & 1.58 & theorem\\
stream (bin.)         & 18.8 & 10.8 & 1.58 & theorem\\
whole-file            & 32.5 & 24.6 & 4.41 & theorem\\
lrat\_isa (bin.)      & 0.76 & 0.72 & 0.03 & verdict\\
cake\_lpr             & 12.7 & 8.26 & 8.93 & verdict\\
lrat-check            & 2.84 & 2.77 & 0.04 & verdict\\
\addlinespace
\multicolumn{5}{@{}l@{}}{$R(B_3,B_7) \le 20$: 58.7 GB; wall in min}\\
stream                & 54   & 3,110 & 4.87 & theorem\\
stream, solver pipe\textsuperscript{a} & 347 & 5,170 & 4.85 & theorem\\
cake\_lpr             & 43 & 2,560 & 12.9 & verdict\\
lrat-check (stdin)    & 10.8 & 631  & 0.62 & verdict\\
\bottomrule
\end{tabular}
\par\smallskip\raggedright
\textsuperscript{a}The check pipelined with the solve (\Cref{sec:evalHex}); wall is
that of both together; the solve alone took 294~min.
\end{table}

\subsection{The empty hexagon at scale}\label{sec:evalHex}

The empty-hexagon theorem states that every set of 30 points in
the plane in general position contains six points forming an empty
convex hexagon. \citet{HeuleScheucher24} established the theorem
with SAT, and \citet{Subercaseaux24} formalized this SAT approach
with cake\_lpr's verdicts recorded as an axiom (\Cref{sec:modes}).
We re-solved the problem over the CNF that the formalization's
encoder produces for $n = 30$. We imported all the certificates the
solvers produced: 174~TB across 312,418 cubes. This ran without any
certificate failures. We processed the cubes in 4,882 groups, each
on one cluster node as one solver-to-checker pipeline, without
writing any certificate to a file. This experiment
is not an end-to-end pipeline run in the sense of \Cref{sec:trust}.
We rely on the original formalization's encoding, which remains
trusted for this case. The import was run on an earlier version of
our tool that did not have the per-import feed constants of
\Cref{sec:streammode}. Therefore, all its native-evaluation axioms mention one feed constant, and axioms about one
constant that refute different leaves are not jointly satisfiable.
We do not claim that theorem. Rebuilding the imports with per-import constants would require re-solving
all 174~TB.

The certification takes about a quarter of the solving CPU time on
1,400 groups of the re-solve. We chose these groups by position,
not cost, and so their leaves are smaller than average. We can
project this to the whole re-solve process and conservatively
estimate 21\%. The ``stream, solver pipe'' row of \Cref{tab:tools} reports wall-clock time, CPU time, and the memory of the Lean process while the solver writes the 58.7~GB certificate into a pipe and the Lean process checks it simultaneously. The wall-clock time is 18\% above the
solve alone, since a full pipe blocks the solver. The memory per group reflects the largest single
leaf in the group, since each leaf restarts the checker
(\Cref{fig:scatter}).

\subsection{Boundary compaction}\label{sec:evalK}

Recall from \Cref{sec:compaction} that boundary compaction removes
the holes of the clause array at a boundary, in an interval of $K$
feed blocks. We measured this operation in stream mode on the two
large certificates of \Cref{tab:compact} and on one leaf. For the certificate of $R(B_2,B_{13}) \le 29$ of size 491~GB, the
compaction reduces the largest memory requirement from 23.0 to
3.67~GB. It requires 1.45 times the CPU time. However, the uncompacted memory requirement grows with
the certificate, so a 2.7~TB certificate of this family would need about 130~GB
and a 27~TB one more than 811~GB.
Consequently, with an enforced 4.3~GB memory limit, only the
compacted import of the 58.7~GB certificate goes through.
\av{On that certificate, the first compaction removes the holes accumulated so far, and later compactions remove the holes of one interval each. }%
We varied $K$ from 16 to 1024 and observed that the memory
requirement stays stable between 2.16 and 2.28~GB over the entire
range, whereas the CPU time grows with the number of compactions,
as each compaction takes a few seconds. On leaf certificates, where the checker restarts on each leaf, no
holes accumulate. Hence, compaction saves almost no memory but costs
time in the renumberer. On the largest queen-domination leaf of
16.8~GB, the largest amount of memory this process required reduced only from 2.34 to 1.80~GB,
while the wall time almost doubled, from 12 to 23 minutes. We
conclude that compaction pays off on long certificates, where
many holes accumulate, but not on leaves.

\begin{table}
\caption{Boundary compaction on the 58.7~GB certificate of
\Cref{tab:tools} and on the 491~GB certificate $R(B_2,B_{13}) \le
29$ of the same family, stream mode; $K$ is the compaction
interval, and the two limit rows ran under an enforced 4.3~GB
memory limit.}
\label{tab:compact}
\footnotesize
\setlength{\tabcolsep}{4pt}
\renewcommand{\arraystretch}{1.05}
\begin{tabular}{@{}lrrrl@{}}
\toprule
certificate & $K$ & compactions & CPU (s) & RSS (GB)\\
\midrule
58.7 GB, 4.3 GB limit & none & 0 & \multicolumn{2}{l}{killed at 33 min}\\
58.7 GB, 4.3 GB limit & 256 & 166 & 3,450 & 2.18\\
58.7 GB & none & 0 & 3,110 & 4.87\\
58.7 GB & 1024 & 41 & 3,000 & 2.28\\
58.7 GB & 64 & 667 & 5,130 & 2.16\\
58.7 GB & 16 & 2,668 & 11,300 & 2.17\\
\addlinespace
491 GB & none & 0 & 22,900 & 23.0\\
491 GB & 256 & 879 & 33,200 & 3.67\\
\bottomrule
\end{tabular}
\end{table}

\subsection{Fault injection and resumption}\label{sec:evalT2}

To empirically verify that a fault cannot yield a theorem, as we
argued in \Cref{sec:checker,sec:streaming}, we injected ten faults
into a six-chunk file-mode import of PHP(10,9). These were every mutation of a snapshot or of a chunk's text that we could construct. For
comparison, we also ran a clean build and a resumed build
(\Cref{tab:injections}\cv{ in \Cref{app:displays}}). Indeed, every fault stopped the build with no theorem, each as a step check evaluating to \lean|false| at the first module that mentions the mutated object. The truncated snapshot was parsed without a sign of a problem, but the result of the parse was rejected by the step check of the next chunk. We also injected three faults into the stream of a PHP(9,8) import:
a truncated certificate, some garbage bytes, and a killed producer. Each injection resulted in the feed returning no block. Thus, the check returned \lean|false| and the module did not compile. If we substitute an earlier
snapshot of the same import for a later one, we get a rejection
from the step check, although every clause in it was indeed derived in the same import.
\av{This is due to the fact that the step check compares the state after the chunk, serialized, for equality with the output snapshot that was named in the statement (\Cref{sec:chaining}). }%
If we resume the build with the artifacts of
four out of six chunks deleted, then Lake rebuilds exactly the four
deleted chunks, each starting from its stored input snapshot. This
produces the identical theorem with the identical six
native-evaluation axioms.

\subsection{End-to-end case studies}\label{sec:evalT3}

We ran five case studies (\Cref{tab:cases}) through the end-to-end
pipeline (\Cref{sec:trust}), consisting of (i) a mathematical
statement in Lean, (ii) an encoder written in Lean, (iii) a proof
that unsatisfiability of the encoding implies the statement, and
(iv) a cube-and-conquer import (\Cref{sec:cover}). The theorems of the \emph{Ramsey
number} and the \emph{van der Waerden numbers} are stated directly
and are the mathematical statements. The theorems of \emph{queen
domination} are stated over the parsed file, with a proof of containment. For the \emph{Keller graph} problem, the import refutes the encoder's term, an encoding using symmetry breaking, where a Lean
proof shows that the symmetry-breaking argument is correct and the
refutation implies the statement about the graph. The Ramsey,
queen-domination, and Keller case studies used streamed imports from
stored certificates. We also recomputed the axiom dependencies of
every completed theorem from the compiled modules. Indeed, they
contain exactly the three standard axioms and one native-evaluation
axiom per evaluation (\Cref{sec:modes}).
\begin{table*}
\caption{The end-to-end case studies. ``Against'' = how the
theorem is related to the encoder's term: stated against it; the parsed base
proved contained in it; or through the graph-level implication.
Solve and import are CPU totals over the row's leaves; the import
CPU of queen domination is summed from per-module build times. ``Certificate'' = bytes the solvers produced; ``retained''
= certificate bytes on any filesystem after the import (zstd-compressed where marked zst; excl.\ the cover certificates). ``std'' = Lean's three
standard axioms; beyond one native axiom per leaf, each row has one
for the cover certificate, one for the check that the leaf modules
cover the parsed cubes, and one for the lower-bound witness where the theorem states one (queen domination: two containment checks instead; $w(2;3,18)$: additionally one per re-cubed sub-leaf and one per sub-cover).}
\label{tab:cases}
\scriptsize
\setlength{\tabcolsep}{\casesSep}
\renewcommand{\arraystretch}{1.15}
\begin{tabular}{@{}>{\raggedright\arraybackslash\tightlines}p{\casesA}>{\raggedright\arraybackslash\tightlines}p{\casesB}>{\raggedright\arraybackslash\tightlines}p{\casesC}rrrr>{\raggedright\arraybackslash\tightlines}p{\casesD}>{\raggedright\arraybackslash\tightlines}p{\casesE}@{}}
\toprule
case study & theorem & against & leaves & solve CPU & certificate & import CPU & retained & axioms\\
\midrule
\textbf{Ramsey} $R(4,4) = 18$ & \lean|r44|: every 2-coloring of $K_{18}$ has a monochromatic $K_4$, and some 2-coloring of $K_{17}$ has none & term & 1,024 & 6.48 h & 85.5 GB & 3.26 h & 85.5 GB & std + 1,027 native\\
\textbf{queen domination} $n = 19$ & \lean|qdom19|: every dominating set of at most 10 queens on the $19 \times 19$ board is a symmetry image of one of 11 listed & contained & 262,144 & 34.1 h & 566 GB & 21.9 h & 95.9 GB (zst) & std + 262,148 native\\
\textbf{Keller graph} $G_{7,3}$ & \lean|keller_graph_7_3|: $G_{7,3}$ has no clique of size $2^7$ & term, graph implication & 21,557\textsuperscript{a} & 50.6 h & 773 GB & 199 h & 204 GB (zst) & std + 21,710 native\\
\textbf{van der Waerden} $w(2;3,17) = 279$ & \lean|w23_17|: every 2-coloring of $\{1..279\}$ has a monochromatic 3-term progression in the first color or 17-term progression in the second, and some 2-coloring of $\{1..278\}$ has neither & term & 770,893 & 224 h & 2.78 TB & 111 h & 0 & std + 770,896 native\\
\textbf{van der Waerden} $w(2;3,18) = 312$ & \lean|w23_18|: as above for 18 and 312, 311 & term & 909,558 & 4,045 h & 25.9 TB & 1,087 h & 0 & std + 914,425 native\\
\bottomrule
\end{tabular}
\par\smallskip\raggedright
\textsuperscript{a}Plus the 148 pieces of the derivation chain (\Cref{sec:deriv}) and the Keller paragraph's three small imports.
\end{table*}
\injectionstable

\paragraph{Ramsey number.}
The theorem \lean|r44| states $R(4,4) = 18$ \citep{Greenwood55},
that is, every two-coloring of the edges of $K_{18}$ contains a
monochromatic $K_4$, and 18 is the smallest such order. The
minimality is witnessed by the Paley graph on 17 vertices; what the
solver needs to show is the first part, that no two-coloring of
$K_{18}$ avoids a monochromatic $K_4$. The encoder
\lean|Ramsey.encode| produces a formula over edge variables. The
import alone took half of the solve CPU, about 140 CPU-seconds per
GB of certificate, which includes the building of modules
(about 53 CPU-seconds per GB for the checker alone, from \Cref{tab:tools}).

\paragraph{Queen domination.}
The theorem \lean|qdom19| states that every set of at most 10 queens
attacking every square of the $19 \times 19$ board is a symmetry
image of one of 11 listed placements. \lean|qdom13| to \lean|qdom18|
state the same for smaller board sizes. The class-count theorems
certify that the given placements are pairwise inequivalent.
\citet{RostamiBright25} computed these numbers by SAT and corrected 43 to 371 for $n = 16$ (OEIS sequence A002563 \citep{OEIS-A002563} still lists 43). We confirm this with our theorem. We
re-solved these instances on our cluster with an unmodified CaDiCaL
that produces LRAT, without the original computation's trusted
solver-added clauses. The solved files drop some tautological
clauses of the encoding. Hence, the parsed file is proved to be
contained in the encoder's clause list, together with the blocking
clauses that block the listed solutions. The import took about two
thirds of the solve CPU, again about 140 CPU-seconds per GB.

\paragraph{Keller graph.}
The vertices of the Keller graph $G_{7,3}$ are the words of length
7 over $\{0, \ldots, 5\}$. Two vertices are adjacent when the
corresponding words differ by exactly 3 in some coordinate and
differ in at least two coordinates. \citet{Keller20} resolved the
last open dimension of Keller's cube-tiling conjecture by refuting
a clique of size $2^7$ in this graph using SAT with symmetry
breaking. We formalized the symmetry-breaking argument for all
$d \ge 5$, $s \ge 2$ using around 2,000 lines of Lean code, no
Mathlib. We proceeded by induction over $d$, starting from three
small imports in dimensions 4, 5, and 6. We did not formalize the
reduction from cliques to tilings. There exists a formalization in
Lean~3 \citep{Clune23}. The symmetry-breaking clauses of the original computation
give rise to our certified derivation (\Cref{sec:deriv}), which we
import as a chain of 148 pieces. We solved the cubes against the
derived formula. The chain required 180 of the 199 CPU-hours of the
import.

\paragraph{Van der Waerden numbers.}
The theorems \lean|w23_17| and \lean|w23_18| state $w(2;3,17) =
279$ and $w(2;3,18) = 312$, respectively. Here $w(2;3,k)$ is the
least $n$ with the following property: whenever we color the
numbers $1, \ldots, n$ with red and blue, there are three red
numbers in arithmetic progression or $k$ blue numbers in arithmetic
progression. \citet{Ahmed10} computed both values; the second one
took 13.6 CPU-years. His colorings provide witnesses for the lower
bounds. For both imports, the cover certificates, 127~MB and 175~MB, respectively, were streamed as well. The import of $w(2;3,17)$ took about half of the solve CPU,
about 144 CPU-seconds per GB. Nineteen of the 909,558 leaves of
$w(2;3,18)$ exceeded the 24-hour limit per leaf. These 19 leaves
were all in the last one percent of the leaves. We re-cubed these
leaves at depth 8, each into 256 sub-cubes with a sub-cover of its
own, and composed each with Lemma~17 a second time (\Cref{sec:cover}).
The import of $w(2;3,18)$ required about a quarter of the solve CPU, around
151 CPU-seconds per GB.

Re-cubing the leaves until each certificate fits into an in-memory
import is not an alternative. \lean|lrat_proof| runs out of memory
at 63~MB, and the whole-file import requires about 16~GB of memory
per GB of certificate (\Cref{tab:tools}) in each process run by
Lake in parallel. However, the leaf sizes in our case studies show
a long tail: the median leaf of queen domination is 38~kB and its
largest 16.8~GB; the median of $w(2;3,17)$ is 0.3~MB and its
largest 7.4~GB. Re-cubing may also result in an impractically large
number of leaves and, in consequence, a large number of modules and
native-evaluation axioms. For a certificate produced by a third
party, like the 58.7 and 491~GB refutations of \Cref{tab:compact},
we cannot re-cube at all.
\av{Similarly, we cannot re-cube a certified derivation, for instance in our Keller case study. }%
With our
streamed import, we use the leaves as the cubing tool produced
them, since the memory requirement depends on the live clause set.

\section{Conclusion}\label{sec:conclusion}                        %

We demonstrated that a SAT solver's certificate can become a named
Lean theorem while the solver is still running, without a single
byte written to disk. As there is no stored data, the theorem
cannot be rechecked from it. However, the proof assistant records
the mathematical statement itself, not just a verdict about a CNF
formula, and can name, compose, and reuse it.

Our Lean library consists of about 2,300 lines. It makes Lean core's
verified LRAT checker resumable. It adds the stream mode, the file mode, and boundary compaction, and it composes cube-and-conquer cases
into one theorem about the original formula. We did not modify any
line of the core checker. However, we added a small byte reader to the trusted base, next to native evaluation.
We used this library in five end-to-end case studies and the
re-solve of the 174~TB empty-hexagon problem (\Cref{sec:eval}).

Although we focus on Lean in this paper, we think that our core
ideas are general enough to be transferred to other proof
assistants. Our construction requires that the proof assistant can
run compiled code whose input comes from outside the proof, as in
Lean's native evaluation. We expect
that a similar division applies to other certificate formats and to
certificates written by several solver processes in parallel.

\av{\section*{Acknowledgments}
Research was supported by the Austrian Science Fund (FWF) within the
projects 10.55776/COE12 and 10.55776/P36688.}

\cv{\bibliographystyle{ACM-Reference-Format}}
\av{\bibliographystyle{plainnat}}
\bibliography{refs}

\clearpage

\end{document}